# Silicon modulator exceeding 110 GHz using tunable time-frequency equalization

**Hengsong Yue,[1] Jianbin Fu,[2] Hengwei Zhang,[1] Bo Xiong,[1] Shilong Pan,[2] and Tao Chu[1],***

[1]*College of Information Science and Electronic Engineering, Zhejiang University, Hangzhou 310027, China*

[2]*National Key Laboratory of Microwave Photonics, Nanjing University of Aeronautics and Astronautics, Nanjing 210016, China*

*\* Corresponding author: chutao@zju.edu.cn*

**Abstract:** Silicon modulators have garnered considerable attention owing to their potential applications in high-density integration and high-speed modulation. However, they are increasingly challenged by the limited 3 dB bandwidth as the demand for modulation speed in optical communications continues to rise, impeding their ability to compete with modulators made of thin-film lithium niobate. This bandwidth limitation arises because of the parasitic resistance and capacitance in the PN junction of the silicon modulators. This study demonstrates the first silicon modulator exceeding 110 GHz without any resonant structure using a tunable time-frequency equalization technique. This substantial breakthrough enables on–off keying modulation at a rate of 140 Gbaud without digital signal processing. These accomplishments represent the highest bandwidth and maximum baud rate achieved without digital signal processing in an all-silicon modulator, reaching the testing limitations of the experimental system. This opens the possibility of attaining modulation rates of up to 200 or even 300 Gbaud by adopting design strategies such as slow light and technologies such as digital signal processing. This advancement extends the speed capabilities of silicon modulators to the level of thin-film lithium niobate modulators, thereby promoting their application in the broader array of fields, such as linear-drive pluggable transceivers.

## 1. INTRODUCTION

Owing to the rapid increase in data traffic, optical telecommunications and interconnections have presented a growing demand for the bandwidth of electro-optic modulators [1–5]. As vital components, electro-optic modulators are employed to convert electrical signals into optical signals to transmit data to optical fibers [6–11]. Silicon photonics, which is compatible with complementary metal–oxide semiconductor (CMOS) processes, has emerged as a highly promising optical system technology owing to its low cost and high-density integration [12–20]. In the early 2000s, researchers began to explore the use of Si as a photonic material. In 2004, an all-silicon modulator with a 3 dB bandwidth exceeding 1 GHz was demonstrated based on a metal-oxide-semiconductor capacitor structure embedded in an optical waveguide [21]. This technology is compatible with CMOS processes. Subsequent research in 2005 demonstrated a compact silicon modulator based on a resonant structure that not only achieved high-speed operation but also significantly reduced the size of the modulator [22]. Subsequently, the introduction of the PN junction and traveling-wave electrode led to a rapid improvement in performance. In 2007, a 3 dB bandwidth of 30 GHz and data transmission rates up to 40 Gbaud were achieved [23], while 2013 witnessed the realization of the silicon Mach-Zehnder modulator (MZM) with 3 dB bandwidth of 27.8 GHz, supporting data rates of 60 Gbaud [24]. In 2018, a new technique involving substrate removal in a silicon MZM was introduced to improve bandwidth [25]. The modulator achieved an extrapolated 3 dB bandwidth of 60 GHz, allowing OOK modulation up to 90 Gbaud.

Recent advancements have highlighted the potential of silicon modulators for ultrahigh-speed applications. In 2020, an ultrahigh-speed silicon microring modulator with a predicted 3 dB bandwidth of 79 GHz supported an OOK modulation of 120 Gbaud [26]. The following year, a silicon microring modulator achieved 128 Gbaud OOK modulation with low power consumption [27]. In 2022, researchers presented a silicon microring modulator with an extended 3 dB bandwidth of 110 GHz and demonstrated high baud rates of up to 120 Gbaud using appropriate signal processing techniques [28]. Another ultra-compact silicon MZM was also demonstrated with a 3 dB bandwidth of 110 GHz and OOK transmission at 112 Gbaud with the help of a slow-light effect [29]. The present strategies to attain such a high 3 dB bandwidth necessitate the development of specialized resonant structures. This introduces a series of problems, including reduced optical bandwidth, increased temperature sensitivity, increased optical loss, and diminished power handling capabilities [30]. The field of silicon modulators is progressively nearing the zenith of its capabilities, particularly in terms of bandwidth and speed. This has driven research into other materials such as lithium niobate [1,31–33], polymer [7,34,35], and plasmonics [2,36–38]. However, these approaches encounter obstacles, such as instability, limited integration capabilities, and incompatibility with CMOS processes. Consequently, it is desirable to extend the performance boundaries of silicon modulators further, particularly to explore their speed potential over a broader optical bandwidth, thereby facilitating their development in a more comprehensive range of applications.

A high-bandwidth silicon MZM exceeding 110 GHz is demonstrated, achieved by employing a tunable time-frequency equalization technique (TFT). This technique incorporates both time and frequency domains, effectively accounting for the relative delay and frequency response in different sections of the silicon modulator. Integrating the TFT with an MZM can be achieved by incorporating multiple modulation regions and extending the traveling wave electrodes, thereby avoiding the introduction of resonant structures that limit the optical bandwidth. The tunability of the TFT can further expand the bandwidth of the modulator. This expansion is attributed to the tunable frequency response of the silicon electro-optical phase shifter, varying with the bias voltage of the PN junction. In the experiments, the employed TFT effectively achieved a silicon modulator with a 3 dB bandwidth exceeding 110 GHz and enabled OOK modulation at 140 Gbaud without DSP. To the best of current knowledge, this represents the highest bandwidth for an all-silicon modulator, as well as the maximum baud rate without DSP. The 3 dB bandwidth can be further extended by leveraging design strategies, such as the utilization of slow light. There exists the potential to achieve remarkable modulation rates of up to 200 or even 300 Gbaud through the synergistic integration of techniques such as DSP. This challenges the conventional understanding of silicon modulators, positioning them to compete effectively with modulators fabricated from materials such as lithium niobate. Consequently, this has fostered the extensive application of silicon photonics across diverse domains.

## 2. RESULTS

### A. Principle

The configuration of the proposed silicon modulator, which incorporates the tunable TFT, is shown in Fig. 1(a). Electro-optic modulation in the silicon modulator is primarily achieved through the plasma dispersion effect within PN junctions. It operates in a series push-pull configuration, wherein two PN junctions are connected in series. In this configuration, two junction capacitors are connected in series, which halves the capacitance, thereby enhancing the bandwidth of the silicon modulator. The modulator folds the traveling wave electrode and optical waveguide to enhance the compactness and facilitate the extension of the traveling wave electrode. Specifically, it was divided into four regions: a main modulation region, a forward modulation region extending along the main modulation region, a velocity mismatch region extending relative to the curved optical waveguide to introduce an electrical delay, and a reverse modulation region of the same length as the forward modulation region. The reverse modulation region is achieved by implementing inverse doping, which transforms the push-pull configuration from the NPN-type doping employed in the main and forward modulation regions to PNP-type doping. These doping methods have similar operational modes. The two PN junctions in the series push-pull configuration equally share the applied voltage, but they cause opposite phase shifts in the optical signals within the two arms of the silicon modulator. The optical signal first passes through the NPN configuration and then through the PNP configuration. The phase shifts of the optical signals in the two arms of the modulator are always opposite. However, the total phase shift of the optical signal in a single arm is the sum of the phase shifts introduced by the NPN and PNP configurations.

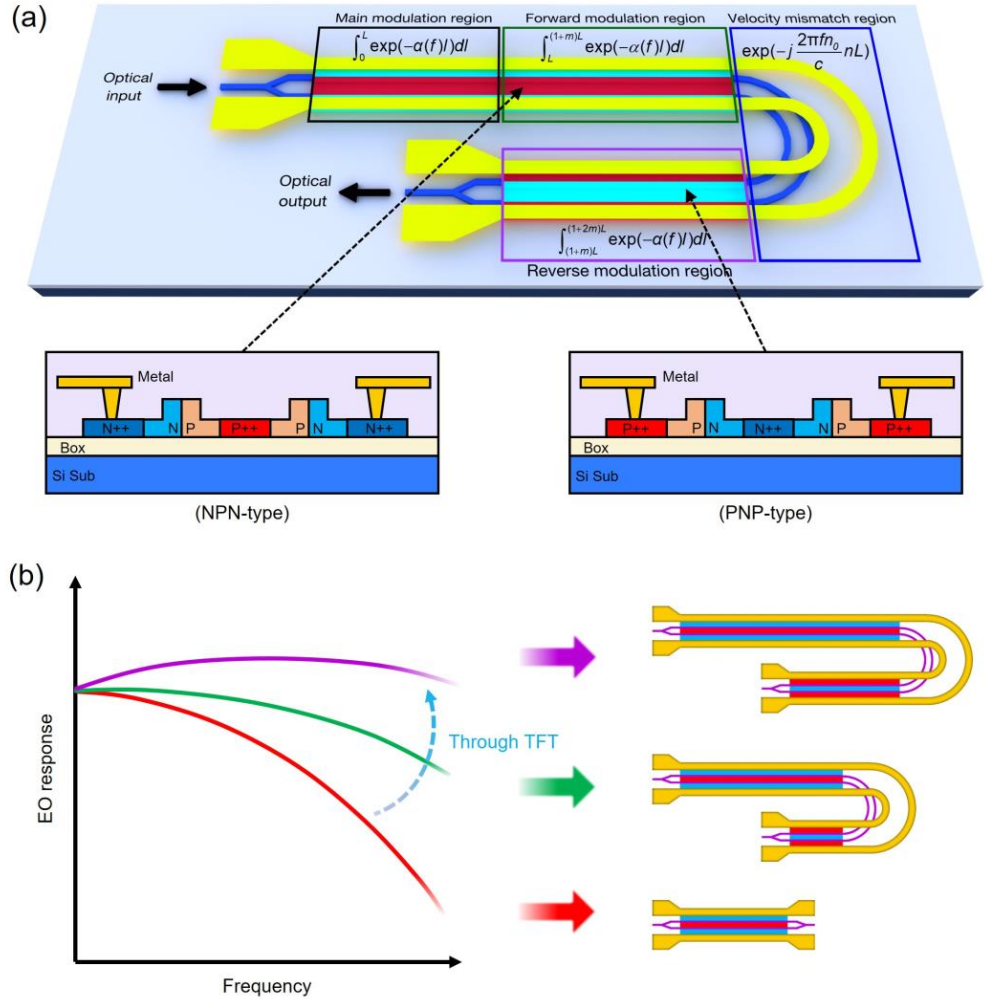

Fig. 1. Principle of the proposed silicon modulator using the tunable TFT. (a) Configuration of the proposed silicon modulator. The modulator consists of a main modulation region, a forward modulation region, a velocity mismatch region, and a reverse modulation region. It features two PN junctions connected in series, operating in a series push-pull configuration. Both the main modulation region and the forward modulation region utilize NPN-type doping, while the reverse modulation region employs PNP-type doping. The response equations for each region are displayed in the figure. (b) The relationship between the amplification of the EO response due to TFT and the modulator structure. Longer forward and reverse modulation regions result in an improved EO response.

All four regions—the main modulation region, forward modulation region, velocity mismatch region, and reverse modulation region—are interconnected through a shared traveling-wave electrode. A single microwave signal applied to this traveling-wave electrode is sufficient to operate the modulator. This signal propagates along the traveling-wave electrode, sequentially passing through the main modulation region, forward modulation region, velocity mismatch region, and reverse modulation region. Similarly, the optical signal traverses these regions in the same sequence within the optical waveguide. It is assumed that the microwave and optical signals propagate at the same velocity.

Within the velocity mismatch region, the lengths of the traveling-wave electrodes and optical waveguides vary, causing different propagation times for the microwave and optical signals. This results in a temporal delay between the forward and reverse modulation regions, which is caused by the velocity mismatch region, leading to the intended mismatch. Synchronous propagation between the microwave and optical signals is maintained throughout the other regions, with no delays introduced, except in the velocity mismatch region. This exception is crucial for affecting the EO bandwidth of the modulator.

Under direct current (DC) signal conditions, the voltage on the traveling-wave electrode remains constant, so the temporal delay does not affect the modulation effect. Since the lengths of the forward and reverse modulation regions are equal and their doping is inverse, the phase shifts of the optical signals induced under a unit voltage are equal in magnitude but opposite in direction. These phase shifts cancel each other out, resulting in complete cancellation. Consequently, the modulator's response is primarily attributed to the main modulation region.

Under microwave signal conditions, the temporal delay introduces a frequency-dependent additional phase to the microwave signal reaching the reverse modulation region. This additional phase, being frequency-dependent, is imparted onto the optical signal as the microwave signal modulates the light in the reverse modulation region. This means that the phase shifts of the optical signals induced by the forward and reverse modulation regions cannot be directly subtracted. Instead, they are combined vectorially, resulting in a composite phase shift stronger than mere subtraction. Thus, the modulator's response is enhanced beyond that of the main modulation region. At certain frequencies where the additional phase reaches $\pi$, the phase shifts from the two regions can be directly summed, resulting in constructive interference. The degree of frequency correlation of the additional phase is determined by the magnitude of the temporal delay, which also governs the frequency of constructive interference. Additionally, considering the continuous attenuation of the microwave signal on the traveling-wave electrode, the response of the forward modulation region is more potent than that of the reverse modulation region. A residual response remains when the phase shifts from the two regions are directly subtracted, regardless of the temporal delay. These effects lead to a new trend in the frequency response curve, overcoming the limitations imposed by microwave attenuation on the modulator's bandwidth.

It is assumed that the velocity and impedance of the silicon modulator are perfectly matched and that the modulation and reverse modulation regions possess identical microwave loss coefficients and modulation efficiencies. The frequency response of the modulator is influenced solely by microwave losses under these assumptions. The length of the main modulation region is denoted as $L$. The length of the forward modulation region is expressed in $mL$, where $m$ represents the relative length ratio of this region to the main modulation region. Similarly, the length of the extended electrodes in the velocity mismatch region is represented by $nL$, with $n$ indicating the length ratio of the extended electrodes in the velocity mismatch region to the main modulation region. In this context, a microwave signal propagating along a traveling-wave electrode can be considered to have the following proportionality:

$$V(l) = V_{in} \exp(-\alpha(f)l) \quad \quad \quad (1)$$

where $\alpha$ is the microwave loss coefficient and $l$ represents the position of the microwave signal within the traveling-wave electrode. Moreover, $V_{in}$ is the amplitude of driving voltage and $f$ denotes the frequency of the input signal. Assuming small-signal modulation, where the response of the Mach–Zehnder interferometer structure is proportional to that of the phase shifter, the response of the three modulation regions in the MZM at frequency $f$ can be expressed as

$$R_{main}(f) \propto \frac{1}{V_\pi L} \int_0^L V(l)dl$$
$$= \frac{V_{in}}{V_\pi L} \int_0^L \exp(-\alpha(f)l)dl \quad (2)$$
$$= \frac{V_{in}}{V_\pi L} \frac{1-\exp(-\alpha(f)L)}{\alpha(f)}$$

$$R_{forward}(f) \propto \frac{1}{V_\pi L} \int_L^{(1+m)L} V(l)dl$$
$$= \frac{V_{in}}{V_\pi L} \int_L^{(1+m)L} \exp(-\alpha(f)l)dl \quad (3)$$
$$= \frac{V_{in}}{V_\pi L} \frac{\exp(-\alpha(f)L)-\exp(-\alpha(f)(1+m)L)}{\alpha(f)}$$

$$R_{reverse}(f) \propto \frac{1}{V_\pi L} \int_{(1+m)L}^{(1+2m)L} V(l)dl$$
$$= \frac{V_{in}}{V_\pi L} \int_{(1+m)L}^{(1+2m)L} \exp(-\alpha(f)l)dl \quad (4)$$
$$= \frac{V_{in}}{V_\pi L} \frac{\exp(-\alpha(f)(1+m)L)-\exp(-\alpha(f)(1+2m)L)}{\alpha(f)}$$

where $V_\pi$ represents the half-wave voltage of the modulator. These terms represent the response of the main modulation region, forward modulation region, and reverse modulation region, respectively. The additional phase caused by the velocity mismatch region is directly proportional to the length of the velocity mismatch region $nL$:

$$P_{mismatch}(f) = \exp(-j\frac{2\pi f n_0}{c} nL) \quad (5)$$

where $n_0$ is the microwave refractive index of the extended electrode. The four items mentioned above are labeled at the corresponding locations in Fig. 1(a). The temporal delay caused by the velocity mismatch region can be expressed as

$$\tau = \frac{n_0 nL}{c} \quad (6)$$

The frequency of constructive interference, where the additional phase reaches π, is given by:

$$f = \frac{c}{2n_0 nL} \quad (7)$$

The specific constructive interference frequency is determined by the modulator's structural parameters, the microwave refractive index $n_0$, and the length of the velocity mismatch region $nL$. These parameters are fixed post-fabrication, thereby fixing the constructive interference frequency as well. As the microwave refractive index $n_0$ and the length of the velocity mismatch region $nL$ increase, the temporal delay $\tau$ increases, leading to a decrease in the constructive interference frequency $f$. Conversely, when these parameters decrease, $f$ increases. Therefore, the response of the overall modulator can be expressed as

$$R_{all}(f) \propto \frac{1}{V_\pi L}\left[\int_0^L V(l)dl + \int_L^{(1+m)L} V(l)dl - P_{mismatch}\int_{(1+m)L}^{(1+2m)L} V(l)dl\right]$$

$$= \frac{V_{in}}{V_\pi L}\left[\begin{array}{l}\frac{1-\exp(-\alpha(f)L)}{\alpha(f)} + \frac{\exp(-\alpha(f)L)-\exp(-\alpha(f)(1+m)L)}{\alpha(f)} \\ -\frac{\exp(-\alpha(f)(1+m)L)-\exp(-\alpha(f)(1+2m)L)}{\alpha(f)}\exp(-j\frac{2\pi f n_0}{c}nL)\end{array}\right] \quad (8)$$

$$= \frac{V_{in}}{V_\pi L}\left\{\begin{array}{l}\frac{1-\exp(-\alpha(f)L)}{\alpha(f)} \\ +\frac{\exp(-\alpha(f)L)-\exp(-\alpha(f)(1+m)L)}{\alpha(f)}[1-\exp(-\alpha(f)mL)\exp(-j\frac{2\pi f n_0}{c}nL)]\end{array}\right\}$$

When $m$ is not equal to 0, the responses of the forward and reverse modulation regions completely cancel each other out at DC. In contrast, residual responses exist at other frequencies owing to the presence of microwave attenuation in the traveling wave electrode. At certain frequencies, the responses exhibited constructive interference. Time-domain information refers to the corresponding electrical delay, which dictates the frequency of constructive interference. This delay is represented by the length of the electrode in the extended-electrode region. On the other hand, the frequency domain information refers to the microwave attenuation values at given lengths, which establish the specific magnitude of constructive interference at the corresponding frequency. This attenuation is denoted by the length of the forward modulation region and microwave loss coefficient. Therefore, it is necessary to consider both time and frequency domains to achieve an optimal bandwidth. When $m = 0$, there are no forward modulation or reverse modulation regions, rendering the proposed silicon modulator a conventional single-drive push-pull silicon MZM. It can be anticipated that an increase in $m$ leads to a more pronounced amplification of the EO response facilitated by TFT. This trend is depicted in Fig. 1(b). The trend can be explained by the decrease in $\exp(-\alpha(f)mL)$ as $m$ increases. Consequently, in the third line of Equation (8), both the fraction term and the terms within the square brackets in the second part of the curly braces increase, enhancing the overall response of the modulator. Additionally, this trend can be further explained through numerical calculations.

    Numerical calculations of the frequency response for each region in the MZM and the overall MZM were performed based on the aforementioned equations to explore the influences of the time and frequency domains. In the analysis, metal losses in the traveling-wave electrode were neglected, focusing only on the microwave losses caused by the shunt conductance from the PN junction. This is because the shunt conductance from the PN junction dominates the microwave losses at high frequencies [39,40]. For silicon modulators, the microwave loss coefficient $\alpha$ caused by the shunt conductance induced by the PN junction is directly

proportional to the square of the frequency. When *m* = 0, the 3 dB bandwidth of the modulator is assumed to be 70 GHz.

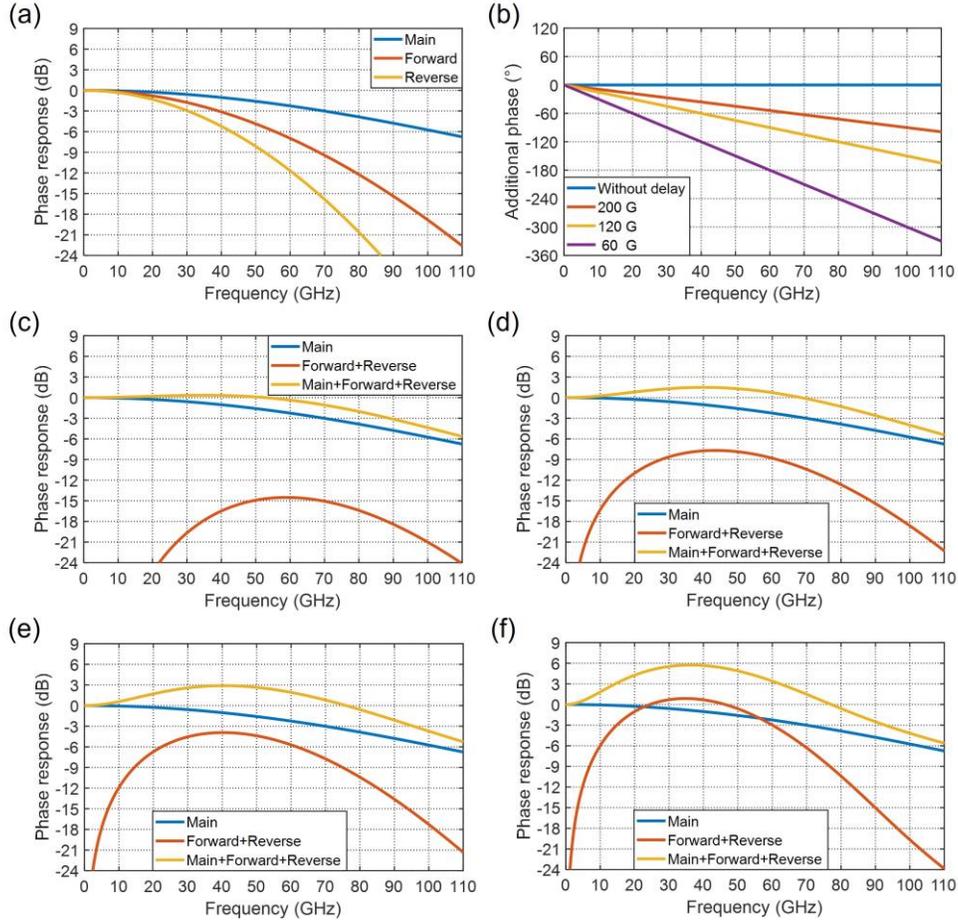

Fig. 2. Phase response analysis of different modulation regions when *m* = 1. (a) The calculated phase response of the three modulation regions in the MZM. (b) The relationship between additional phase and frequency at different constructive interference frequencies. (c) The calculated phase response of the main modulation region, the combined response of the forward and reverse modulation regions, and the combined response of the main, forward, and reverse modulation regions without temporal delay. The calculated phase response of the main modulation region, the combined response of the forward and reverse modulation regions, and the combined response of the main, forward, and reverse modulation regions at constructive interference frequencies of (d) 200 GHz, (e) 120 GHz, and (f) 60 GHz.

The frequency response of each region in the MZM describes the frequency-dependent phase shifts induced in the optical signal by each region, as the regions of the MZM cannot be treated as a singular, complete modulator. The calculated phase response of the three modulation regions for *m* = 1 is shown in Fig. 2(a). The lengths of the three modulation regions are equal, resulting in consistent phase response intensity under DC conditions. The main modulation region has the highest bandwidth, followed by the forward modulation region, and then the reverse modulation region. This is due to the continuous attenuation of the microwave signal as it propagates along the traveling-wave electrode. Figure 2(b) shows the relationship between additional phase and frequency at different constructive interference frequencies. The additional phase in the reverse modulation region reaches 180 degrees at the constructive interference frequency.

The combined response of the forward and reverse modulation regions is obtained by performing a vector subtraction of the response of the reverse modulation region, which includes the additional phase, from the response of the forward modulation region. Further, the combined response of the main, forward, and reverse modulation regions is obtained by performing a vector addition of this result with the response of the main modulation region. Figs. 2(c)–2(f) depict the phase response of the main modulation region, the combined response of the forward and reverse modulation regions, and the combined response of the main, forward, and reverse modulation regions at different constructive interference frequencies. As the constructive interference frequency decreases, the peak value of the combined response of the forward and reverse modulation regions gradually increases, while the frequency corresponding to this peak value gradually decreases. The frequency corresponding to the peak value differs from the constructive interference frequency due to significant attenuation in the response of the forward modulation region at the constructive interference frequency. In contrast, the response of the forward modulation region is stronger at lower frequencies. Notably, the combined response of the main, forward, and reverse modulation regions is equivalent to the overall MZM response. The peak in the combined response of the forward and reverse modulation regions leads to an elevation in the overall MZM response.

Figures 3(a)–3(d) depict the frequency response of the silicon modulator under different time-domain conditions when m is 0.5, 1, 1.5, and 2, and compare these responses with the frequency response of the modulator when $m = 0$. The time-domain effects were illustrated with the constructive interference frequency as a variable to provide a more intuitive representation of the results. It is important to note that displaying the frequency response at different constructive interference frequencies does not imply that these frequencies can be tuned. Once the modulator is fabricated, the temporal delay becomes fixed, and consequently, the constructive interference frequency is also fixed. Figures 3(a)–3(d) are intended to illustrate that both the length ratio of the forward modulation region $m$ and the constructive interference frequency affect the equalization effect. Their effects on the 3 dB bandwidth of the silicon modulator and the bandwidth expansion in comparison to $m = 0$ are detailed in Table 1. A greater value of $m$ indicates a heightened magnitude of the response, consequently leading to a higher 3 dB bandwidth expansion. As the value of $m$ increases, the impact of the constructive interference frequency on the bandwidth expansion gradually diminishes. This phenomenon occurred because a more significant value of $m$ resulted in a longer propagation path for the microwave signal, leading to greater attenuation. Consequently, the response-magnitude amplification effect caused by constructive interference is weakened. In addition, the 3 dB bandwidth expansion exhibits a trend of initially increasing and then decreasing as the constructive interference frequency decreases from 200 GHz to 60 GHz. This observation indicates that an optimal bandwidth expansion exists for each specified value of $m$, which can be attributed to the limited amplification of the response magnitude when $m$ is held constant. Achieving optimal bandwidth expansion requires a constructive interference frequency to match the 3 dB bandwidth of the equalized silicon modulator precisely. It is worth noting that the upward stretching of the response at low frequencies caused by constructive interference may result in a loss of broadband signal quality. It is challenging to avoid such stretching owing to the minimal attenuation of microwave signals at low frequencies.

Using the conditions above, the functional relationship between the optimal bandwidth expansion and variable $m$ was determined. Notably, the behavior of the microwave loss coefficient against frequency in the silicon modulator is influenced by various sources of microwave loss, including metal losses in the traveling-wave electrode and microwave losses caused by the shunt conductance from the PN junction. Metal losses, which are proportional to the square root of frequency, suggest that the actual microwave loss coefficient is not strictly proportional to the square of the frequency but is somewhat lower. Numerous simulations and empirical measurements of EE $S_{21}$ and microwave attenuation indicate that the relationship between the microwave loss coefficient $α$ and frequency is less than quadratic [24,25,41–45]. The frequency response calculations of the proposed silicon modulator, characterized by a microwave loss coefficient proportional to the frequencies raised to the powers of 1 and 1.5, are detailed in the Supplementary Materials Section S1.

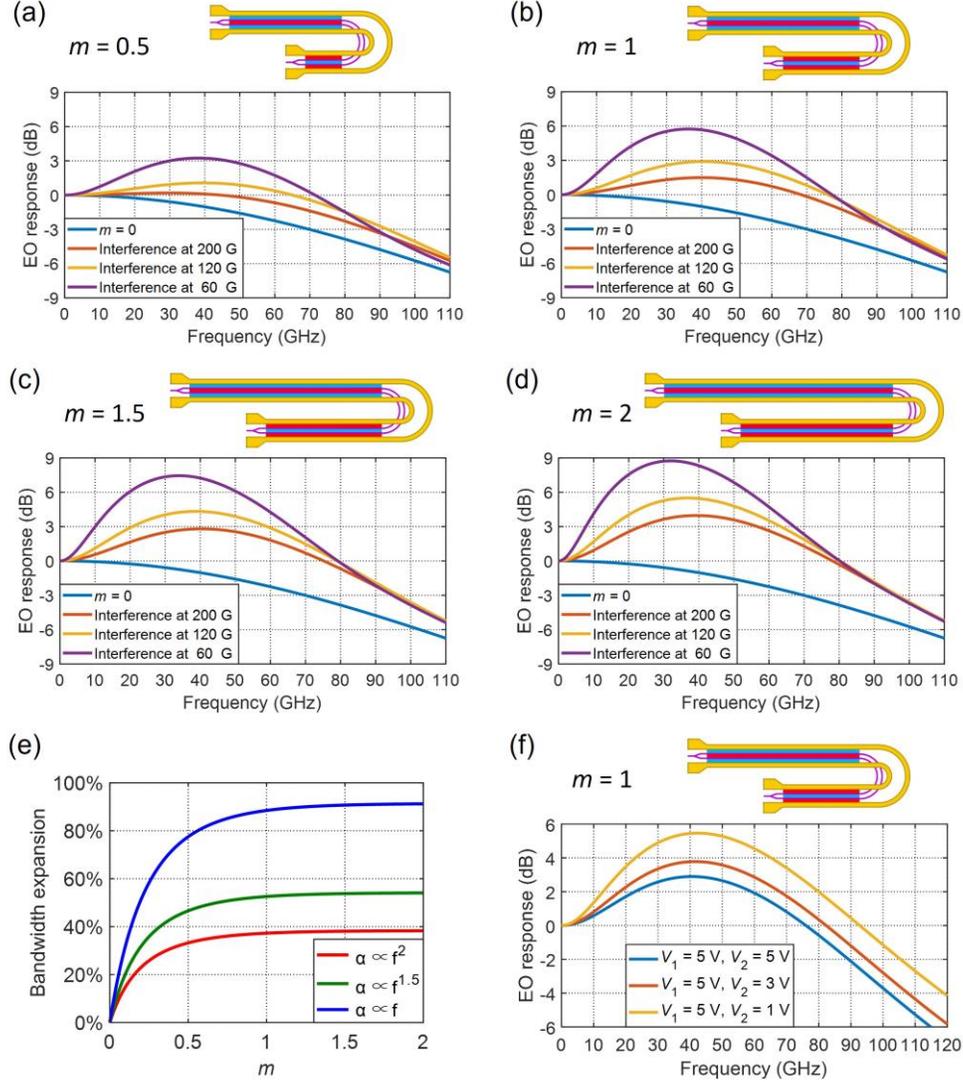

Fig. 3. Frequency response and equalization analysis of the proposed modulator. The calculated frequency response of the proposed silicon modulator with different constructive interference frequencies at (a) $m = 0.5$, (b) $m = 1$, (c) $m = 1.5$, and (d) $m = 2$, respectively, and compare these responses with the frequency response of the modulator when $m = 0$. The velocity and impedance of the silicon modulator are assumed to be perfectly matched. The metal losses in the traveling wave electrode are neglected. (e) The relationship between the optimal bandwidth expansion and $m$. The constructive interference frequency is precisely matched with the 3 dB bandwidth of the equalized silicon modulator. (f) The tunability of the TFT. $V_1$ represents the reverse bias voltage of the main modulation region and the forward modulation region. $V_2$ represents the reverse bias voltage of the reverse modulation region. The variations of microwave loss coefficient and modulation efficiency with reverse bias voltage are deduced from a published paper.

**Table 1. Impact of *m* and Constructive Interference Frequency on Silicon Modulator 3 dB Bandwidth and Expansion.**

| *m* | Constructive Interference Frequency (GHz) | 3 dB Bandwidth (GHz) | Bandwidth Expansion (%) |
|---|---|---|---|
| 0.5 | 200 | 86.9 | 24.1 |
|  | 120 | 92.1 | 31.6 |
|  | 60 | 88.7 | 26.7 |
| 1 | 200 | 92.9 | 32.7 |
|  | 120 | 95.6 | 36.6 |
|  | 60 | 92.8 | 32.6 |
| 1.5 | 200 | 95.1 | 35.9 |
|  | 120 | 96.4 | 37.7 |
|  | 60 | 94.8 | 35.4 |
| 2 | 200 | 96.0 | 37.1 |
|  | 120 | 96.6 | 38.0 |
|  | 60 | 95.8 | 36.9 |

Functional relationships were computed for cases in which the microwave loss coefficient was directly proportional to the 1, 1.5, and 2 powers of the frequency. As illustrated in Fig. 3(e), the optimal bandwidth expansion increases with increasing *m*. This is because the amplification of the response magnitude owing to constructive interference is more significant when *m* increases. However, the rate of increase in the optimal bandwidth expansion decreased with an increase in *m*. When *m* was small, the optimal bandwidth expansion increased rapidly. However, as *m* increases, the rate of increase in the optimal bandwidth expansion gradually decreases and finally reaches a plateau. This is because the microwave signal gradually attenuates along the traveling-wave electrode. Consequently, further amplification of the response magnitude due to an increase in *m* is weakened. Specifically, an increase in *m* indicates that the lengths of both the forward and reverse modulation regions are continuously increasing. The response of the silicon modulator at a given frequency is proportional to the integral of the microwave signal strength along the traveling-wave electrode at that frequency, since all modulation regions share a common traveling-wave electrode. However, this integral reaches a numerical limit despite the increase in the lengths of the modulation regions due to the exponential decay of the microwave signal along the traveling-wave electrode. A limited integral means that the response of the silicon modulator at the corresponding frequency is also constrained, thereby restricting the bandwidth expansion. Consequently, the curves in Fig. 3(e) approach saturation. The respective maximum optimal bandwidth expansions were 91%, 54%, and 38% for powers of 1, 1.5, and 2. The maximum optimal bandwidth expansion increased as the power number decreased. Considering a power of 1.5 is more reasonable for our context (see Section S5 for details). The threshold value for the onset of saturation can be optimized by adjusting the design of the PN junction and the applied bias voltage. These methods can reduce the contribution of the shunt conductance from the PN junction to microwave losses at high frequencies, thereby decreasing the microwave loss coefficient.

The calculated frequency response of the proposed silicon modulator, when the modulation and reverse modulation regions have different reverse bias voltages, is shown in Fig. 3(f). This numerical calculation demonstrates the tunability of the TFT, which occurs when the PN junction bias voltages differ across the modulation regions. This tunability is attributed to the EO modulation characteristics of the PN junction. In silicon traveling-wave MZMs based on PN junctions, a lower reverse bias voltage narrows the depletion region of the PN junction, resulting in a higher equivalent capacitance and, consequently, a greater microwave loss coefficient. Simultaneously, the change in depletion region width per unit voltage increases, leading to higher modulation efficiency. Therefore, decreasing the reverse bias voltage of the reverse modulation region enhances its DC response, thereby suppressing the overall DC response of the MZM. This adjustment expands the modulation bandwidth compared to the scenario where all modulation regions share the same reverse bias voltage. In this calculation, *m* is set to 1, and the constructive interference frequency is established at 120 GHz. The main and forward modulation regions share the same reverse bias voltage ($V_1$), whereas the reverse modulation regions have separate reverse bias voltages ($V_2$). Additionally, the 3 dB bandwidth of the silicon modulator with $m = 0$ was considered to be 70 GHz when the PN junction was reverse-biased at 5 V. The calculation was performed with reference to the measured half-wave voltage and 3 dB bandwidth at different reverse bias voltages provided in a published paper [46]. Variations in the microwave loss coefficient and modulation efficiency can be deduced from these data with respect to the changes in the reverse bias voltage. The curve normalized to the DC power when all modulation regions have the same reverse bias voltage is represented by the blue line in Fig. 3(f). The red and yellow lines in Fig. 3(f) represent the response curves relative to the blue curve when the modulation and reverse modulation regions have different reverse bias voltages. These relative curves highlight the changes in esponse due to variations in the reverse bias voltage of the reverse modulation region.

The response at low frequencies is gradually suppressed as the reverse bias voltage of the reverse modulation region decreases. After normalization in Fig. 3(f), it is shown that a lower reverse bias voltage exhibits a higher high-frequency response. Specifically, when both reverse bias voltages are set at 5 V, the proposed silicon modulator achieves a 3 dB bandwidth of 95.6 GHz, which aligns with the results shown in Fig. 3(b). Reducing the reverse bias voltage in the reverse modulation region to 3 V and 1 V increased the response at high frequencies by 0.92 dB and 2.62 dB, respectively. This results in an extended 3 dB bandwidth of 101.3 GHz and 112.1 GHz. This demonstrates the potential of the TFT to expand the bandwidth further.

## B. Device Characterization

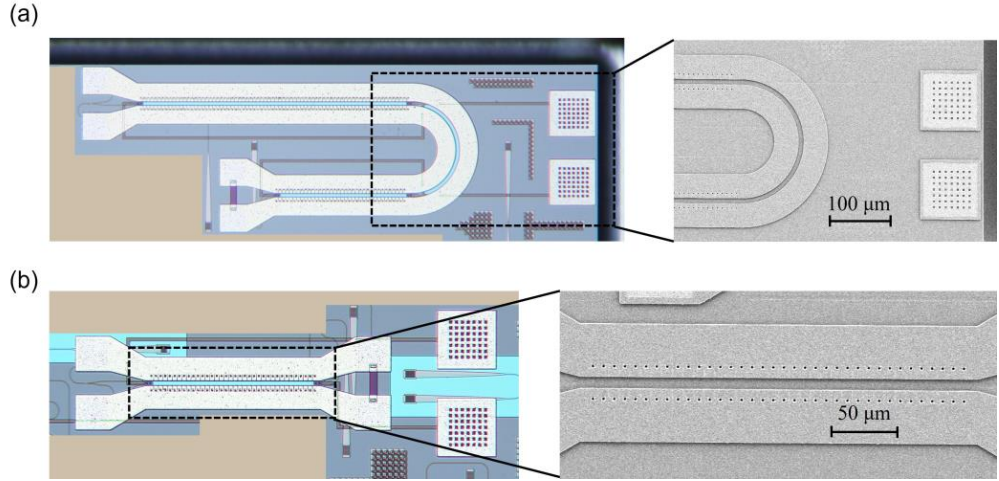

Fig. 4. Micrograph and SEM image of the fabricated silicon modulators. (a) Micrograph and SEM image of the silicon modulator with $m = 1$. The curvature radius of the curved optical waveguide is approximately 100 μm. In contrast, the traveling wave electrode in the velocity mismatch region exhibits an additional length of approximately 77 μm. (b) Micrograph and SEM image of the silicon modulator with $m = 0$. The length of all modulation regions is 0.3 mm. Brown markings obscure distinct devices not relevant to the study.

Silicon modulators are designed to be compatible with standard CMOS processes. They are fabricated using the commercial foundry services provided by Advanced Micro Foundry, whose silicon photonics platform was developed based on standard CMOS technology. The micrograph and scanning electron microscope (SEM) image of the fabricated silicon modulator employing a tunable TFT are displayed in Fig. 4(a). The lengths of the main, forward, and reverse modulation regions were all 0.3 mm, corresponding to a design with $m = 1$. The traveling wave electrode in the velocity mismatch region exhibits an additional length of approximately 77 μm in comparison to the curved optical waveguide. Although the theoretical corresponding constructive interference frequency for this length is extremely high, it can effectively enhance the bandwidth by considering the existence of velocity mismatches in various regions of the silicon modulator. The curvature radius of the curved optical waveguide is approximately 100 μm. An on-chip terminator is fabricated at the end of the traveling wave electrode, possessing a resistance of 50 Ω. Two DC voltages are applied to the two heavily doped regions between the optical waveguides. These are used as the bias voltages for the two modulation regions and the reverse modulation region, respectively. Additionally, a conventional silicon modulator featuring only the main modulation region was also fabricated, corresponding to a design with $m = 0$ (see "Methods" for more details of device design and fabrication), and the micrograph and SEM image are shown in Fig. 4(b).

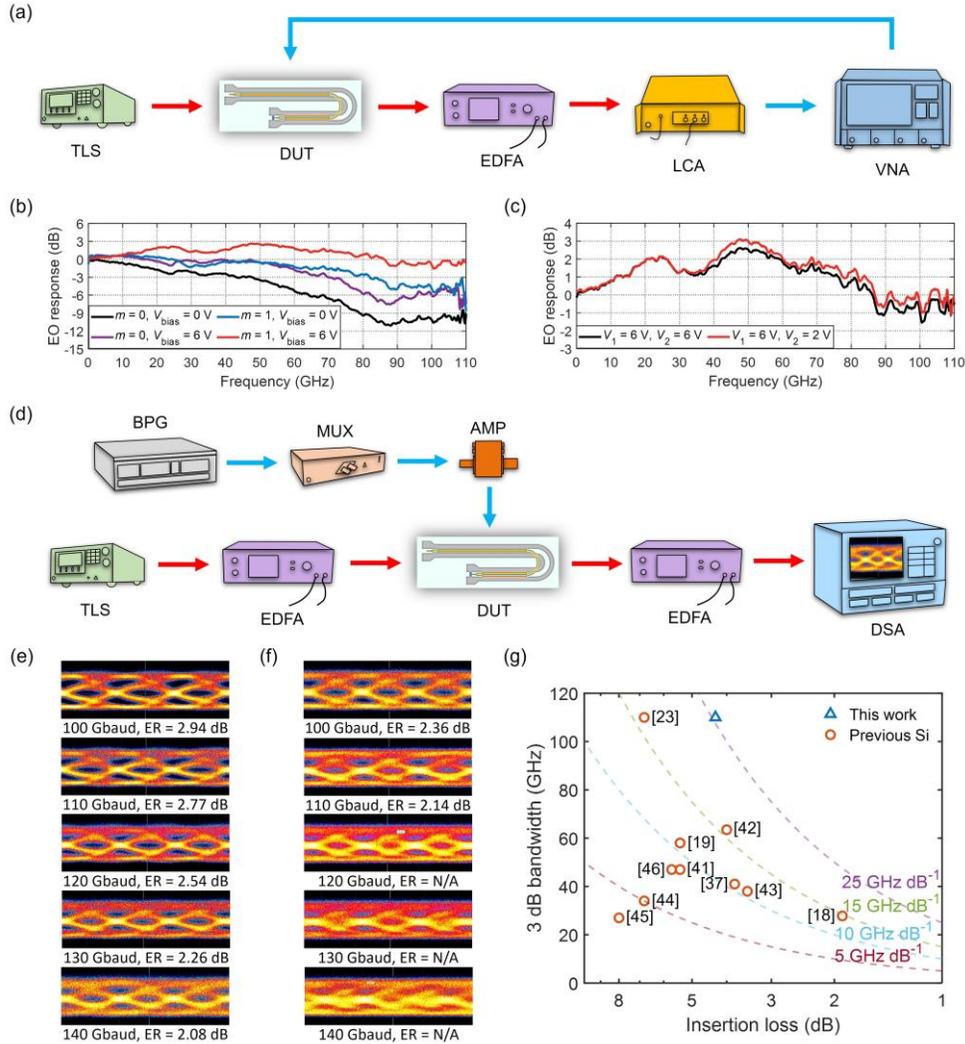

Fig. 5. EO characterization and eye diagram analysis of silicon modulators. (a) Experimental setup for measuring the EO bandwidth of the silicon modulator. (b) EO response

characterization of silicon modulators with $m = 0$ and $m = 1$ when all modulation regions have the same reverse bias voltage. (c) EO response tunability of the silicon modulator with $m = 1$ when modulation regions and reverse modulation regions have different reverse bias voltages. (d) Experimental setup for measuring the OOK eye diagrams of the silicon modulator. (e) The OOK eye diagrams without DSP for the silicon modulator with $m = 1$. (f) The OOK eye diagrams without DSP for the silicon modulator with $m = 0$. N/A, not applicated. (g) Performance comparison among this work and previous silicon modulators, focusing on the trade-off between 3 dB bandwidth and insertion loss. Dashed lines show loss-bandwidth trade-off curves.

The electro-optic (EO) bandwidth measurement was conducted by collecting the functional relationship between the response and frequency under various bias voltages, and the experimental setup is shown in Fig. 5(a) (see "Methods" for more details of the experimental setup). Figure 5(b) shows the measured EO response of the silicon modulators with $m = 0$ and $m = 1$ at reverse bias voltages of 0 and 6 V. All modulation regions of the silicon modulators have the same reverse bias voltage. At a 0 V bias, the 3 dB bandwidth of the silicon modulators with $m = 0$ and $m = 1$ are 43.1 and 81.9 GHz, respectively. At a 6 V reverse bias, the 3 dB bandwidth of the silicon modulators with $m = 0$ and $m = 1$ are 70 GHz and larger than 110 GHz, respectively. To the best of the authors' knowledge, the measured 3 dB bandwidth of 110 GHz represents the highest bandwidth ever documented for an all-silicon modulator. At this point, the response of the silicon modulator with $m = 1$ at 110 GHz is approximately -1 dB. The frequency response of the silicon modulator with $m = 1$ exhibits an initial increasing and then decreasing trend, demonstrating good agreement with numerical calculations.

Figure 5© illustrates the EO response of the silicon modulator with $m = 1$ when the modulation and reverse modulation regions have different reverse bias voltages, thereby reflecting the tunability of the TFT. The black line represents the curve normalized by the power at DC when all modulation regions have the same reverse bias voltage. The red line represents the response curve normalized by the power at DC when the modulation and reverse modulation regions have different reverse bias voltages. The reverse bias voltage in the modulation region is denoted as $V_1$, while the reverse bias voltage in the reverse modulation region is referred to as $V_2$. As $V_2$ decreases from 6 V to 2 V, the response at high frequencies is increased by approximately 0.5 dB. This phenomenon is consistent with the numerical calculations shown in Fig. 3(f). This demonstrates the capacity of the TFT to expand the bandwidth further and even aid in compensating for the signal impairment in the link, thereby enhancing the transmission rate of the signal. It is important to note that the modulation bandwidth is extended by suppressing the low-frequency response of the MZM. However, excessive suppression can cause significant overshoot in the high-frequency response, thereby reducing overall signal quality. The optimization of the modulation bandwidth should be guided by the specific signal quality requirements of the application. Additionally, precisely defining the maximum achievable modulation bandwidth is challenging. The modulation efficiency of the PN junction is significantly higher under forward bias than under reverse bias. Adjusting the bias voltage in the reverse modulation region could reduce the overall DC response of the MZM to zero, complicating the optimization of the bias for maximum bandwidth.

An eye diagram test was conducted to analyze the performance enhancement caused by applying the TFT. Figure 5(d) shows the experimental setup employed to obtain the OOK eye diagrams of the silicon modulator. Two DC power sources are used to generate reverse bias voltages, $V_1$ and $V_2$, which are applied to the silicon modulator with $m = 1$. The silicon modulator with $m = 0$ requires the application of only a single reverse bias voltage (see "Methods" for more details of the experimental setup). The peak-to-peak voltage of the applied signal during our experiments was maintained at 5 V.

Figures 5(e) and 5(f) exhibit the OOK eye diagrams obtained from the silicon modulator with $m = 1$ and $m = 0$, respectively, without any DSP. The bias voltages of these modulators are dynamically tuned in real-time to attain the utmost optimal shape of the obtained eye diagram. For the silicon modulator with $m = 1$, the open eye diagram demonstrates a maximum baud rate of 140 Gbaud, with an extinction ratio (ER) of 2.08 dB. The quality of its eye diagrams is noticeably superior to that of the silicon modulator with $m = 0$. The latter exhibits a maximum baud rate of only 110 Gbaud, and ER is also inferior to that of the silicon modulator with $m = 1$. The ERs of both modulators are constrained by the short lengths of the modulation regions and the modulation efficiencies, which have not been specifically optimized. It should be noted that the maximum achievable baud rate of 140 Gbaud for the silicon modulator with $m = 1$ is primarily constrained by the limited bandwidth of the amplifier, multiplexer, and RF probe in the experimental setup rather than the bandwidth of the silicon modulator itself. Despite these limitations, it remains the highest baud rate for an all-silicon modulator without any DSP. The loss-bandwidth performance of our modulator is compared with that of previous silicon modulators in Fig. 5(g), visually showcasing the superiority of the TFT. Significantly, our modulator outperforms existing silicon modulators, achieving ultra-high bandwidth with relatively low losses. The ultra-high bandwidth of our modulator could facilitate data operations surpassing 200 Gbaud.

## 3. DISCUSSION

**Table 2. Comparison of Several Performance Metrics for Silicon Modulators.**

| Reference | Type | EO Roll-Off | Baud Rate (Gbaud) | Optical Bandwidth (nm) | Loss (dB) | Length (mm) | Vπ·L (V·cm) |
|---|---|---|---|---|---|---|---|
| [25] | MZM | −1.2 dB at 50 GHz | 90 | >50[a] | 5.4 | 2 | 1.4 |
| [29] | MZM | −3 dB at 110 GHz | 112 | ~8 | 6.8 | 0.124 | 0.96 |
| [41] | MZM | −2 dB at 67 GHz | 120 | >50[a] | 6 | 2 | 3 |
| [47] | MZM | −3 dB at 47 GHz | 100 | >50[a] | 5.4 | 2.5 | 1.35 |
| [48] | MZM | −3 dB at 63.5 GHz | / | >50[a] | 4 | 4 | 2.8 |
| [49] | MZM | −3 dB at 38 GHz | 115 | >50[a] | 3.5 | 3 | 3.6 |
| [53] | MZM | / | 112 | >50[a] | 6.9 | 2.47 | 1.5 |
| [26] | MRM | 0 dB at 67 GHz | 120 | <1[a] | / | 0.008[b] | 0.8 |
| [27] | MRM | -3 dB at 77 GHz | 128 | <1[a] | / | 0.006[b] | 0.42 |
| [28] | MRM | −3 dB at 110 GHz | 120 | <1[a] | / | 0.008[b] | 0.8 |
| [54] | MRM | −3 dB at 62 GHz | 120 | <1[a] | / | 0.004[b] | 0.42 |
| This work | MZM | −1 dB at 110 GHz | 140 | >50[a] | 4.3 | 0.9 | 4.86[c] |

[a]Estimated from the modulator type.
[b]The radius of the MRM.
[c]See section S4 for the details.

Table 2 compares the performance of the modulator presented in this work with other modulators demonstrated on silicon. This equalization technique enables our modulator to achieve an EO roll-off of only 1 dB at 110 GHz by considering both the time and frequency domains simultaneously. It exhibits remarkable efficacy in overcoming the inherent bandwidth limitations of silicon modulators. To the best of our knowledge, this is the first time a silicon modulator without a resonant structure has achieved a 3 dB bandwidth of 110 GHz. The elimination of resonant structures in the design preserves the broad optical bandwidth of the Mach-Zehnder interferometer and its low sensitivity to temperature fluctuations, in contrast to alternative high-bandwidth silicon modulators [27–29,54]. This contributes to the optimal utilization of spectral resources and conservation of the power budget, providing an efficient solution for parallel data transmission. The first demonstration of a remarkable 140 Gbaud data transmission experiment, achieved without any DSP, enables a wide range of applications and highlights the potential of silicon photonics in high-speed data transmission. Our MZM exhibits optical bandwidth performance comparable to that of other MZMs, due to its use of a standard Mach-Zehnder interferometer structure. The device experiences moderate insertion loss, which meets the requirements of most applications. Additionally, our modulator offers a distinct advantage in terms of length, being longer only than slow-light MZMs. This compactness facilitates easier integration into densely packed optoelectronic systems, although it still cannot match the smaller size of microring modulators. Our device also demonstrates relatively low modulation efficiency within the range typical for MZMs, generally underperforming compared to microring modulators. The modulation efficiency was calculated based on the results of measuring the dependence of the modulation phase on voltage using a DC voltage (see Section S4 for details). The half-wave voltage can then be estimated from the modulation efficiency and the length of the modulator. Dividing the modulation efficiency by the length of the modulator yields an estimated half-wave voltage of approximately 54 V. In addition, the tunability of the TFT has significant implications for adapting the modulator to various system requirements and optimizing its performance in different scenarios.

It is worth noting that there is a fundamental difference in principle between the equalized silicon modulator and the segmented silicon modulator, despite their similar structures, which allows for the mutual reference of transfer functions [41,55,56]. The segments of the segmented silicon modulator typically have identical doping and similar modulation characteristics, whereas the equalized silicon modulator introduces additional complexity by intentionally inversing the doping in the reverse modulation region compared to the main and forward modulation regions. This difference in doping, coupled with the consideration of the continuous attenuation of the microwave signal as it propagates through the traveling-wave electrode, allows the EO bandwidth of the equalized silicon modulator to be extended. Besides, the equalized silicon modulator requires only a single driving signal, whereas the segmented silicon modulator requires a number of driving signals corresponding to the number of segments. This difference simplifies the driving electronics and can lead to more efficient system integration.

It is important to acknowledge potential areas for improvement in this study. For instance, a traveling wave electrode is an essential component in the modulator design. However, this research did not dedicate specific efforts toward optimization, resulting in a restricted bandwidth in conventional silicon modulators. There is a strong potential to significantly improve the ER and reduce the half-wave voltage by increasing the length of the modulation regions and exploring methods to enhance modulation efficiency through the optimization of materials and processes. This would make the modulators more suitable for practical communication applications. Furthermore, a more precise evaluation and optimization can be achieved by simultaneously considering velocity mismatch, impedance mismatch, and other factors leading to microwave losses. In addition, exploring the compatibility of the TFT with different modulator types and its integration into existing communication networks are valuable avenues for future research. There is the potential for further breakthroughs by combining it with other advanced design strategies, such as leveraging the slow light effect [29]. In addition, the proposed TFT can also be applied to other optoelectronic platforms. The TFT is expected to exhibit superior performance on platforms with lower microwave loss coefficients, owing to the correlation between the frequency response and the microwave loss coefficient. The versatility of this approach provides a promising pathway for innovation and applications in optoelectronic platforms, thereby enabling improved performance and signal quality.

In conclusion, a tunable TFT for silicon modulators was proposed and demonstrated, achieving a 3 dB bandwidth exceeding 110 GHz. This technique can be seamlessly integrated with an MZM by making simple modifications to the optical waveguide and traveling-wave electrode, eliminating the need for resonant structures that impose limitations on the optical bandwidth. In the experiment, the TFT enhanced the 3 dB bandwidth of the silicon modulator from 70 GHz to beyond 110 GHz. Additionally, the baud rate of the OOK modulation increased from 110 to 140 Gbaud without DSP, constrained by the limited bandwidth of the experimental system. These two groundbreaking achievements indicate the potential of silicon photonics technology to facilitate high-speed data transmission in the future. Moreover, this technique provides a strong foundation for innovation and application in other optoelectronic platforms.

## 4. MATERIALS AND METHODS

## A. Device Design and Fabrication

Silicon modulators were fabricated on a silicon-on-insulator wafer possessing a top silicon layer of 220 nm thickness and a buried oxide layer with a thickness of 2 µm, utilizing the fabrication service rendered by Advanced Micro Foundry (Singapore). The width of the traveling wave electrode is 40 µm with a gap of 9 µm. The width of the optical waveguide was 500 nm, and the thickness of the slab was 90 nm. The PN junction was established inside the optical waveguide, and the junction interface was displaced 50 nm from the center of the waveguide towards the N-type region. Heavily doped regions were utilized to establish ohmic contact with the aluminum electrode, maintaining a separation of 750 nm from the center of the optical waveguide. This separation is employed to achieve a balance between optical absorption loss and modulation speed. The closer these heavily doped regions are to the optical waveguide, the higher the optical absorption losses. However, a smaller separation decreases the effective resistance of the PN junction because the heavily doped regions have lower resistivity. This reduction in resistance, combined with the associated capacitance of the modulator, results in higher modulation speeds. Conversely, increasing the separation reduces modulation speed but decreases optical absorption losses. The Mach-Zehnder interferometer features a path length difference between its two arms, allowing the adjustment of the optical carrier wavelength to alter the operating point.

The silicon modulator with $m = 1$ exhibited an insertion loss of 4.3 dB. The silicon modulator with $m = 0$ exhibits an insertion loss of 2 dB. The static ERs are 33.2 dB and 35.3 dB, respectively. The path length difference between the two arms of the MZM results in free spectral ranges of 9.6 nm and 9.3 nm, respectively. The estimated half-wave voltage, based on the modulation efficiency and the length of the modulation region, is 54 V. The propagation loss of the doped silicon waveguide is approximately 4.2 dB/mm, and each grating coupler exhibits an insertion loss of approximately 4.8 dB (see Sections S2, S3, and S4 for details).

## B. Experimental Setup for High-Speed Measurements

EO response measurements were performed using a vector network analyzer (Keysight N5245B) and a 110 GHz lightwave component analyzer (Newkey GOCA-110) [57]. For the eye diagram measurements, an optical carrier was generated using a C-band tunable laser source (Santec TSL550) and further amplified with an erbium-doped fiber amplifier (EDFA) prior to its coupling into the silicon modulator via a grating coupler. Polarization alignment was achieved using a polarization controller. Two serial data streams were generated using a bit pattern generator (SHF 12104A) and directed towards a 128 Gb/s 2:1 multiplexer (SHF C603 B), resulting in the generation of an electrical OOK signal with a doubled baud rate. This signal was then amplified using a 66 GHz amplifier (SHF M804 B) with an output Vpp of 5 V. It was subsequently fed to the silicon modulator through a 40 GHz RF probe. Subsequently, the modulated light at the output of the modulator is amplified by an additional EDFA and connected to a digital serial analyzer (DSA) for analysis. OOK eye diagrams were captured using DSA without employing any DSP.

**Acknowledgements.** The authors would like to extend sincere thanks to Jianhong Liang, Zongxin Xu, Xu Tan, and Shuang Wang for their valuable support during the testing phase of this work.

**Data Availability statement.** Data supporting the findings of this study are available from the corresponding author upon request.

**Disclosures.** The authors declare no conflicts of interest.